\documentclass[aps,twocolumn,showpacs,floatfix,nofootinbib,preprintnumbers,superscriptaddress,amsmath,amssymb]{revtex4-1}

\usepackage{graphicx}
\usepackage{epsfig}					
\usepackage{bm}
\usepackage{color}
\usepackage{float}
\usepackage{dcolumn}
\usepackage{subfigure}
\usepackage{multirow} 
\usepackage{amsmath}

\renewcommand{\figurename}{Fig.}

\newcommand{\citefig}[1]{\figurename~\ref{#1}}
\newcommand{\citeeq}[1]{Eq.~(\ref{#1})}

\newcommand{\MFPT}[1]{\langle \tau\left( {#1} \right) \rangle}

\begin{document}

\title{Direct Coupling of Free Diffusion Models to Microscopic Models of Confined Crystal Growth and Dissolution}

\author{J{\o}rgen H{\o}gberget} 
\affiliation{Department of Physics, University of Oslo, N-0316 Oslo, Norway}

\author{Dag K. Dysthe}
\affiliation{Department of Physics, University of Oslo, N-0316 Oslo, Norway}

\author{Espen Jettestuen} 
\affiliation{IRIS AS, P.O. Box 8046, N-4068 Stavanger, Norway}
\affiliation{Department of Physics, University of Oslo, N-0316 Oslo, Norway}

\begin{abstract} 
We couple a free solute diffusion model to 
a model of crystal surface growth represented by, 
but not limited to, a (2 + 1)-dimensional solid-on-solid (SOS) model confined by a flat surface.
We use kinetic Monte Carlo (KMC) with dissolution rates based on nearest-neighbor interactions
to solve the Master equation for the surface dynamics, and we use an offlattice random walk 
to model the Fickian diffusion of the solute particles.
The two solvers are coupled directly through deposition rates of the free particles
calculated using the mean first passage time (MFPT) of deposition that is found to scale as $r^{-4}$. 
Two variants are studied: ignoring (radial) and not ignoring the line of sight (pathfinding).
Reference models such as uniform concentration (random deposition) 
and lattice diffusion (crystal lattice extended into the liquid) are used for comparison.
We find that the macroscopic limit of the surface dynamics is reproduced by all models.
The free diffusion models produce a lower equilibrium roughness and a smaller height autocorrelation length than the reference models,
and are found to behave very well in tight confinements. It is also demonstrated that lattice diffusion does not work well in tight confinements. 
The two MFPT models behave very similarly close to equilibrium and for dissolution, but becomes increasingly different with increasing surface growth speed. 
The model is put to use by simulating a cavity with a flux boundary condition at one side.
The conclusion is that the new model excels in confinement, 
and line of sight can in practice be ignored since the dominant deposition sites likely are in line of sight, 
which minimizes the CPU-time needed in the coupling.
\end{abstract}


\maketitle

\section{Introduction}
\label{sec:Intro}

Surface growth and dissolution of a solid from solution requires mass
transport by diffusion to and from the liquid-solid interface. When the
liquid is confined to nanometer thick films the diffusion in
confinement becomes an important control on the process. Moreover, when
confined, liquids behave very differently from their respective bulk
phases~\cite{Chiavazzo2014, Diallo2015}.  Feedback between
growth and dissolution, mass transport, and confinement induced stresses
are known to lead to instabilities that reduce diffusion paths and
increase dissolution rates by many orders of
magnitude~\cite{DenBrok1998,Dysthe2002}.


Molecular Dynamics (MD) is well suited for simulating transport in nanoconfined liquids~\cite{Camara2004, Kalcher2013, Sun2013, Ori2014, Chiavazzo2014, Yasuoka2015}.
However, since MD operates on the time scale of atomic vibrations,
it suffers from a time scale problem when faced with growth and dissolution processes,
which involve time scales that are orders of magnitude larger.
This is the reason why kinetic Monte Carlo (KMC) is widely used in these cases~\cite{Voter2007}.  

The challenge is that KMC cannot be directly coupled to separate dynamic processes 
unless these dynamics are either given directly in terms of transition rates,
or they do not affect the KMC transition rates. 
For instance, free diffusion impacts
the KMC state transition rates since it sets the deposition rates. 
Hence KMC is unable to produce a time step until a deposition rate is supplied, but 
the free diffusion is unable to integrate in time until a time step is suppled.

For surfaces in open reservoirs this problem has been solved by 
splitting the system into two: one pure solute part which can be iterated in time using practically any algorithm, 
and one part consisting of the crystal surface and a minimal amount of solute which is iterated in time using rate based lattice diffusion,
with a shared concentration and/or flux boundary between the two subsystems that is iteratively solved for consistency at all times~\cite{Drews2004,Pricer2002,Pricer2002a,Tello2004}.



In our confined system the restriction of having particles diffusing on a lattice 
could induce unwanted effects because the ability of the solute to adapt to its surroundings
is potentially very important. 
We will therefore in this paper construct an algorithm that couples free diffusion to the KMC surface dynamics,
and benchmark it against a lattice diffusion implementation.

%

Since the crystal surface is dynamic, and acts as a spatial boundary condition to the liquid, 
using a continuum description of the liquid is not practical. These continuum methods 
also scales badly with increasing dimensionality~\cite{Plapp2000a}.
Continuum models also fail to incorporate important particle fluctuations,
which leaves a discrete particle based model optimal~\cite{VanZon2005}.
Hence we will base our algorithm on a discrete particle model.

The paper is structured as follows: In Sec.~\ref{sec:model} we briefly introduce the surface model before we 
give an in-depth derivation of the free diffusion model and the coupling to the surface dynamics.
Following in Sec.~\ref{sec:Results} we present a detailed analysis of the model 
where we compare its behavior to that of lattice diffusion and a uniform concentration model.
Concluding the results section we apply the derived model to simulate a system where the 
lattice diffusion and the uniform concentration model are not well suited.  
In Sec.~\ref{sec:Discussions} we summarize and conclude the paper.

\section{Model}
\label{sec:model}

\begin{figure*}
 \begin{center}
  \includegraphics[width=0.3\textwidth]{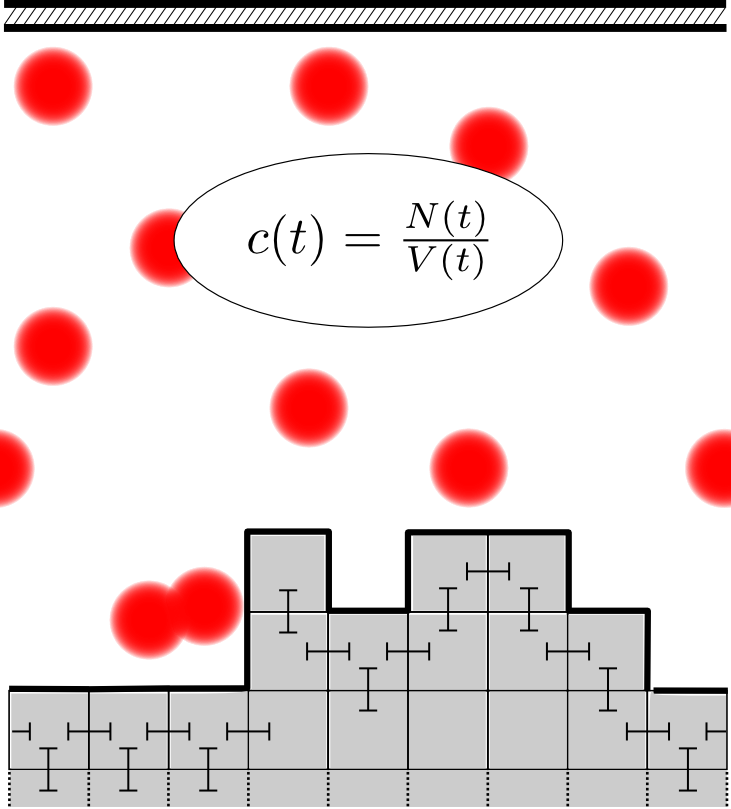}
  \hspace{0.5cm}
  \includegraphics[width=0.3\textwidth]{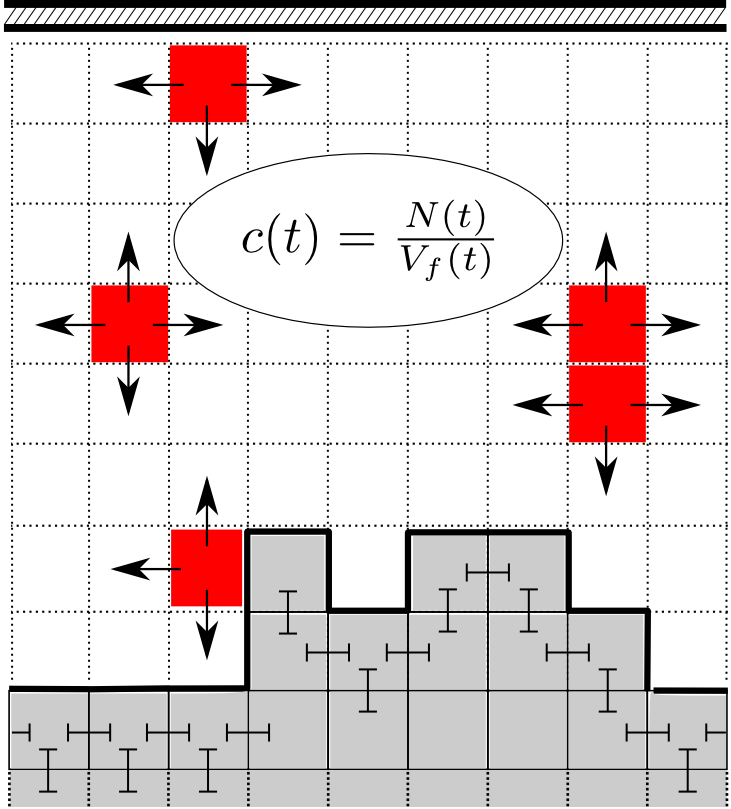}
  \hspace{0.5cm}
  \includegraphics[width=0.3\textwidth]{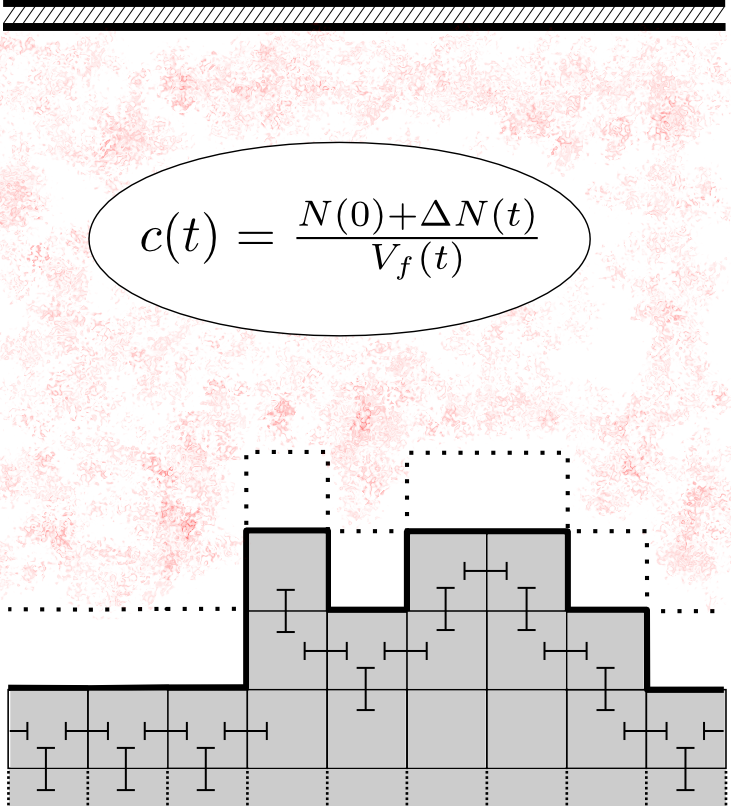}
 \caption{Illustrations of the different diffusion models and the surface model used in this paper.
 To the left we have the free diffusion model we derive and analyze,
 and the middle and right figures are reference models known as 
 lattice diffusion and uniform concentration, respectively.
 The solid-on-solid (SOS) surface is indicated by a thick solid line,
 and the flat confining surface is on the top of each respective part.
 Particle-particle bonds are shown as solid capped lines. Since the liquid is ideal there are no such bonds between solute particles. The center equation 
 shows how the concentration as a function of time $c(t)$ is represented in the diffusion model, with $N(t)$ being the number of solute particles, $V(t)$ is the cavity volume,
 and $\Delta N(t)$ is the change in solute particles since $t=0$. 
 The models in the middle and to the right uses a different volume
 to calculate the concentration, 
 $V_f(t) \equiv V(t) - Al_0$, where $Al_0$ is the volume of one surface layer,
 since the sites immediately above the surface (stippled lines in the right figure)
 are guaranteed not to contain any particles.}
\label{fig:diffusion_models}
 \end{center}
\end{figure*}

\subsection{Surface}

%
%

The surface is modeled using a (2 + 1)-dimensional solid-on-solid (SOS) surface with
nearest neighbor interactions. Above the surface there is an inert flat surface
placed at a fixed position such that the initial average separation is $\Delta h$.

The allowed surface transitions are dissolution and deposition at top sites only (the SOS restriction), i.e.~
processes which either reduces or increases the local height by one unit.
The exception is if there is no space around a site to dissolve to,
or no space between the surfaces to deposit.

The dissolution rate is~\cite{Gilmer1972}

\begin{equation}
\label{eq:rate}
 \mathcal{R}(p) = \nu\exp(-\alpha n(p)),
\end{equation}

\noindent
where $n(p)$ is the number of nearest neighbors particle $p$ has, 
and $\alpha \equiv E_b/kT$, with $E_b$ being the nearest neighbor bond energy, 
$k$ being the Boltzmann constant, 
and $T$ being the temperature.

The deposition rate will depend on how we model the diffusion in the liquid surrounding the surface.

\subsection{Liquid}

In the liquid, we will assume the dynamics to be governed by Fickian diffusion, that is, the particle concentration obeys the diffusion equation

\begin{equation}
\label{eq:diffusion}
\frac{\partial c(\mathbf{r}, t)}{\partial t} = D\nabla^2 c(\mathbf{r}, t),
\end{equation}

\noindent
where $c(\mathbf{r}, t)$ is the concentration at a time $t$ at the position $\mathbf{r}$ and $D$ is the diffusion coefficient,
here assumed to be independent of position. 

We model the liquid as ideal, 
such that $E_b = 0$ for solute particles.
Inserting this into Eq.~(\ref{eq:rate}) yields a constant transition rate $\mathcal{R}(p_s) = \nu$ for solute particles. 
The diffusion coefficient for solute particles on a lattice with separation $l_0$ is thus

\begin{equation}
\label{eq:D}
 D = \mathcal{R}(p_s) l_0^2 = \nu l_0^2,
\end{equation}

\noindent
where $l_0$ is the lattice separation. 
This is the diffusion coefficient we will use throughout this work.

This is a simplification since, in general, the value of $\nu$ in the liquid will differ from that of the surface.
Hence $D$ should be a free parameter of the liquid. Nevertheless,
we will use Eq.~(\ref{eq:D}) here. 
Using a specific value for $D$ (in units of $\nu$) should be straight forward.


%
%
%

\subsubsection{Free diffusion}

Free diffusion is included by not constraining the solute particles to a lattice.
The solute particles do not interact with anything except the surfaces which act as physical boundaries (hard-sphere interactions).

In order to move the solute particles in accordance with the diffusion equation in \citeeq{eq:diffusion}, we update their
positions according to the recurrence relation~\cite{Plapp2000a}

\begin{equation}
\label{eq:free_diff_recurrence}
 x(p)_{i+1} = x(p)_i + r_N
\end{equation}

\noindent
where $x(p)_i$ represents a Cartesian degree of freedom for solute particle $p$ at a time step $i$, and $r_N$ is a normal distributed random number with
variance $2D\delta t$, with $\delta t$ being the time step. If a suggested position is illegal (i.e.~blocked by the surfaces), we redraw a new position.
Alternatively one could reject the move or reflect the move.

The concentration is calculated using the cavity volume $V(t)$ as 

\begin{equation}
 c(t) = \frac{N(t)}{V(t)},
\end{equation}

\noindent
where $N(t)$ is the number of solute particles at time $t$.

\subsubsection{Deposition rates}

For the standard liquid models used in surface growth such as the reference models we use (see Appendices~\ref{app:lattice} and~\ref{app:uniform} for details),
the deposition rates are immediately available as either a constant 
in the case of a uniform concentration model, 
or as an implicit transition from a lattice cell in the liquid to one on the surface.

In the case of free diffusion this kind of approach is not possible, 
since the particle dynamics are not given in terms of rates.
We cannot simply let the particles diffuse and attach if they collide with the surface either,
since we in kinetic Monte Carlo do not a priori know the time the surface stays in its current state unless
we a priori know the deposition rates. This creates a circular dependency between the time step and the deposition rates.
To break this circular dependency we will introduce a model for the solute particle deposition rates. 

Given that a particle $p$ is located at some position $\textbf{r}_p$,
we want to calculate the deposition rate at a specific surface site $i$.
This could be measured as the average time it takes separate diffusion processes 
starting out at $\textbf{r}_p$ to reach the surface at site $i$ for the first time.
This quantity is referred to as the mean first passage time (MFPT) of deposition. 
If we denote the MFPT of deposition as $\MFPT{\delta r_{ki}}$, the deposition rate becomes

\begin{equation}
\label{eq:mfpt_deprate_orig}
 \mathcal{R}_+(p, i) \propto 1/\MFPT{\delta r_{ki}},
\end{equation}

\noindent
where $\delta r_{ki}$ is a characteristic distance between solute particle $p$ at a position position $\textbf{r}_p$ and surface site $i$. 

Calculating the MFPT is expensive. The reason we use kinetic Monte Carlo for the surface dynamics 
is to achieve larger time scales, hence it would render this choice obsolete if 
the individual cycles became too slow. We therefore aim here to derive a generic form 
of the MFPT expression using what we know from the diffusion equation in Eq.~(\ref{eq:diffusion}).

Comparing the dimensions of $\MFPT{\delta r}$ to those of the parameters in Eq.~(\ref{eq:diffusion})
suggests a relationship on the form

\begin{equation}
\MFPT{\delta r} \propto \delta r^2/D.
\end{equation}

\noindent
This expression is spherically symmetric, and a
deposition at a specific site $i$ can very crudely be thought of as escaping this shell
through a specific square hole of area $l_0^2$. 
A very crude correction for this is to weigh the time 
by the ratio of the area of the hole to that of the full shell $4\pi r_{ki}^2$, 
which yields

\begin{equation}
 \MFPT{\delta r_{ki}} \propto \frac{4\pi r_{ki}^2}{l_0^2} \frac{r_{ki}^2}{D}.
\end{equation}

\noindent
Inserting this back into Eq.~(\ref{eq:mfpt_deprate_orig}) yields the deposition rate

\begin{equation}
\label{eq:mfpt_deprate_final}
 \mathcal{R}(k, i) = \kappa D/r_{ki}^4, 
\end{equation}

\noindent
where the value of $\kappa$ is obtained by requiring that on average the system should
equilibrate at the same concentration as the uniform concentration model.

This expression has the same form as what is calculated for a narrow escape through a hole in 
a sphere in Ref~\cite{Oshanin2010}.


We will model the characteristic distance $r_{ki}$ in Eq.~(\ref{eq:mfpt_deprate_final}) 
in two different ways: one using radial distances, 
and another using the shortest path, which can be efficiently calculated using the A$^\ast$ algorithm~\cite{Hart1968}.
The idea is that the latter takes into account physical obstacles along the way, whereas the former does not.
We will refer to these methods as radial and pathfinding MFPT, respectively.

In \citefig{fig:ovitofig} an example of a shortest path calculation is shown, where
a particle finds its way to a deposition site on the other side of an obstacle. 
A radial calculation here would over estimate the probability of reaching this site.

\begin{figure}
	\begin{center}
		\includegraphics[width=0.45\textwidth]{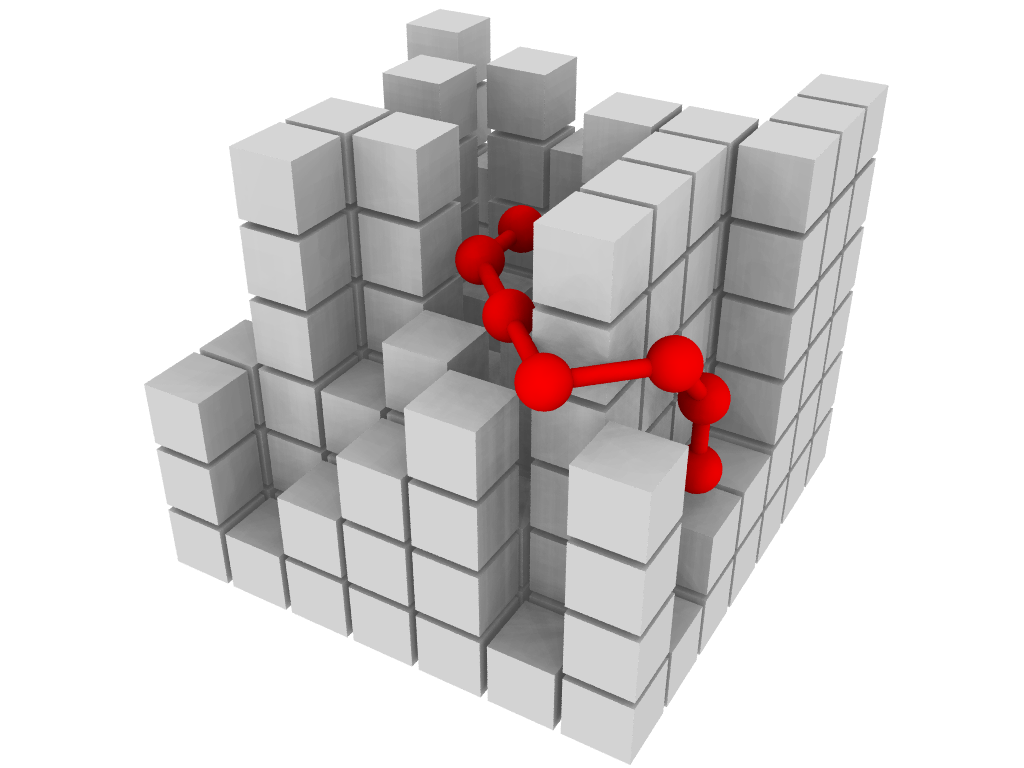}
	\caption{An illustration of how the shortest path to a surface site is found in the pathfinding MFPT model.
	A particle (here present in the center of the surface) follows a shortest path to the deposition site 
	on the other side of the obstacle. Each node in the pathfinding is marked by a sphere 
	and the transition between nodes are indicated by a thick line.
 Since the diffusion in the liquid is free, a node is connected to all 26 neighboring nodes.
 Every surface site except the destination is blocked, since arriving here would cause a bond to the surface. 
 	The image has been rendered using Ovito~\cite{Stukowski2009a}.}
	\label{fig:ovitofig}
	\end{center}
\end{figure}

\subsubsection{Implementation details}

Following is the algorithm used to couple the kinetic Monte Carlo dynamics to the free diffusion process: 

\begin{enumerate}
 \item Calculate all transition rates $\mathcal{R}_i$ using Eq.~(\ref{eq:rate}) for dissolution reactions and Eq.~(\ref{eq:mfpt_deprate_final}) for the deposition reactions.
 \item Calculate the time step $\delta t = -\log(r)/\mathcal{R}_\mathrm{tot}$, where $r$ is a uniformly distributed random number and $\mathcal{R}_\mathrm{tot}$ is the sum of all rates
 calculated in the first step.
 \item Assign a probability $\mathcal{P}_i = \mathcal{R}_i/\mathcal{R}_\mathrm{tot}$ to each transition, draw one at random, and execute it.
 \item Diffuse all solute particles not involved in the transition for a time $\delta t$ (this may need to be split into shorter intervals).
\end{enumerate}

\noindent
We have set the maximum allowed time step for a single free diffusion step equal to $0.01$.

Since the continuous coordinate solute particles are allowed 
close to the discrete surface, problems might occur where e.g.~a solute particle becomes 
stuck in a surface hole without being able to select a legal diffusion step. 
To avoid this we have simply chosen to give all solute particles a wiggle-room of $l_0/2$ to each side.

Since the $r^{-4}$ term in the MFPT decays very quickly with distance $r$, 
we calculate the deposition rate only inside a cube centered on the diffusing particle of lengths $l=2r_\mathrm{max}+1$,
where we set $r_\mathrm{max}=3$. 
This is referred to as the maximum distance method, 
and is thoroughly analyzed in Ref.~\cite{Schwarz2013}, 
where it is stated that $\sqrt{6}r_\mathrm{max}\in [7,9]$ is safe for all practical purposes ($3\sqrt{6}\simeq 7.35$ in our case).
This also means that particle with a vertical position greater than the maximum surface height plus $r_\mathrm{max}$
has a deposition rate equal to 0.
Hence the CPU time required to calculate the 
deposition rate only scales with the number of particles close to the surface, and not the 
total number of particles unless the volume is very small.

\section{Simulations and Results}
\label{sec:Results}

For the simulations we have used the rejection free kinetic Monte Carlo (KMC) algorithm~\cite{Voter2007}.
First we will find the values of $\kappa$ in the mean first passage time (MFPT) from
\citeeq{eq:mfpt_deprate_final} by requiring the system to on average 
equilibrate at the same concentration as the uniform concentration model from Appendix~\ref{app:uniform}. 
Once $\kappa$ is obtained we will compare the free diffusion model and the reference models to the analytical 
solutions to the macroscopic limit derived in Appendix~\ref{app:analytical}. 
As a final comparison with the reference models we calculate the equilibrium roughnesses and height autocorrelation lengths for each model.
We then apply the radial model to a system where the other models are not suited in order to demonstrate its necessity.
The selected system is a cavity geometry with a high flux of particles into the system at one boundary.

We implement constant cavity volume by keeping the average separation between the crystal and the confining surface constant
by repositioning the latter.
In simulations done at a constant number of particles, whenever a particle is added or removed by the simulations,
we remove or insert a particle at random, respectively. All simulations have an initial average height $h(0) = 0$,
which means that the position of the confining surface $h_l$ represents the initial surface-surface separation $\Delta h$ as well.

For simulations where the confining surface was fixed, 
the system is initialized in a steady state at $t=0$ for $h(0)$ and $\Omega(0)$
generated by simulating a given number of steps under a constant cavity volume as described in the previous paragraph.

All separations are measured as center-center distances, such that the edge-edge distance is one lattice unit less.

\subsection{Equilibration}
\label{ssec:equilibration_free}

\begin{figure}[t]
	\begin{center}
	\includegraphics[width=0.45\textwidth]{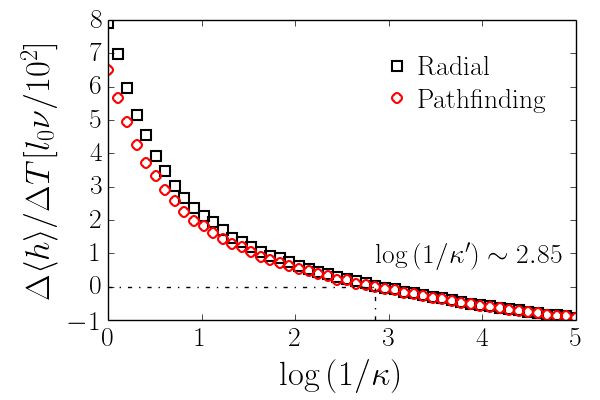} 
	\caption{The growth speed as a function of $\kappa$ for the radial MFPT algorithm (squares) and the pathfinding MFPT algortihm (circles),
	using $\alpha=1$, a constant average surface-surface separation of $\Delta h/l_0 = 10$, 
	and a constant number of particles set such that the concentration equals
	that of the equilibrium concentration of the unconfined uniform concentration model from \citeeq{eq:c_eq_analytical}. 
	The results are obtained by averaging 20 simulations per data point.
	We see that for the less the system grows, the better the two models match, which is expected since these models only affect
	the growth mechanisms.
	The dashed lines are visual aids to see the zero point and the resulting value of $\kappa=\kappa' \sim \exp(-2.85) \sim 0.058$.}
	\label{fig:finding_c}
	\end{center}
\end{figure}
	
\begin{figure}[t]
	\begin{center}
	\includegraphics[width=0.45\textwidth]{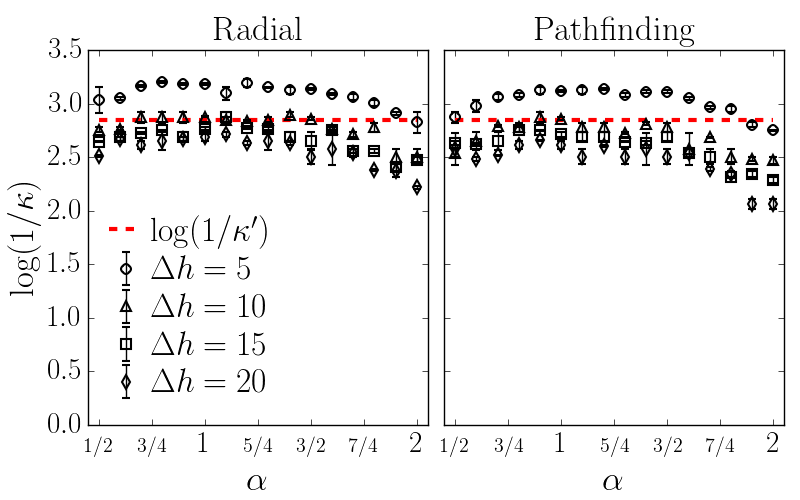} 
	\caption{The value of $\kappa$ as a function of $\alpha$ for various surface-surface separations $\Delta h$
	using the radial (left) and pathfinding (right) MFPT models. 
	These resuls are obtained by bisection of the mean growth speed on the interval $\log\left(1/\kappa\right)\in[1.5, 3.5]$. 
	The error bars indicate the resulting bisection interval, which in the case of large bars indicate that
	the bisection has not converged properly due to statistical noise. 
	The stippled line is a visual aid and denotes the value obtained for $\alpha=1$ and $\Delta h/l_0 = 10$ from \citefig{fig:finding_c}.}
	\label{fig:finding_c_bisection}
	\end{center}
\end{figure}

In order to obtain reasonable values for $\kappa$ in \citeeq{eq:mfpt_deprate_final} for both the
radial and the pathfinding mean first-passage time (MFPT) models, we fix the number of particles and the volume,
and set the initial concentration to the known equilibrium concentration of the 
unconfined uniform concentration model from \citeeq{eq:c_eq_analytical}. 
Since the two diffusion models are very different in nature, this is a crude approximation, 
but it serves as a reasonable way to obtain a fair value for the parameter nonetheless.

We then apply two different methods for estimating the value of $\kappa$ which leads to zero net growth.
In the first method we measure the growth speed as a function of $\kappa$ averaged over several simulations
and interpolate the point where the growth speed is zero, 
and in the second approach we estimate the root of the growth speed as a function of $\kappa$ using bisection.  
The first approach is very accurate, but requires many simulations and much manual labor.
The second approach is completely automated, however, has limited precision due to statistical fluctuations in the growth speed measurements.


The results of the first approach for the radial and pathfinding MFPT
models are shown in \citefig{fig:finding_c}, where we see that for dissolving systems ($\Delta \langle h\rangle /\Delta t < 0$),
the two models behave equally, whereas for growing systems, the two models
diverge more and more as the growth speed increase. This makes sense since for 
dissolving systems, the details of the growth mechanisms should be less important,
and opposite for growing systems. It also makes sense that we see the radial method
producing all around higher growth speeds, since it does not care about lines of sight
and should therefore overestimate the deposition rates. Around equilibrium, however, the two models
are very similar and yield a unified value of $\kappa \sim \exp(-2.85) \sim 0.058$.

The results for bisection for different surface-surface separations $\Delta h$ and $\alpha$ is given in 
\citefig{fig:finding_c_bisection}. Ideally we would want $\kappa$ to be independent of $\Delta h$ and $\alpha$,
however, as we can see this is not the case. One problem is that the assumption that the equilibrium concentration 
is given by \citeeq{eq:c_eq_analytical} is increasingly invalid with decreasing $\Delta h$, since this equation is
derived without accounting for finite-size effects ($\Delta h\to\infty$). This effect is prominent in \citefig{fig:finding_c_bisection}
where the $\Delta h/l_0=5$ results appears shifted in comparison to the other data. 
Moreover, as $\alpha$ increase, particles far away
from a deposition site can contribute significantly to its total deposition probability. 
Our expression for this deposition rate is
very basic and issues in this limit is therefore expected. This effect is also prominent in the figure
where we see the results curve off for higher $\alpha$. These facts does not imply that the model is invalid
in these regimes, it means only that we must expect $\kappa$ to have dependencies on the model parameters,
and ensure that we use a value which is appropriate for the given set.
Nevertheless, the dependency of $\kappa$ on the model parameters appears to be smooth
and well behaved.


%

\subsection{Comparisons with the macroscopic limit}
\label{ssec:sim_vs_analytical}

\newcommand{\scaleone}{0.24}
\begin{figure*}[t]
 \begin{center}
   \includegraphics[width=\scaleone\textwidth]{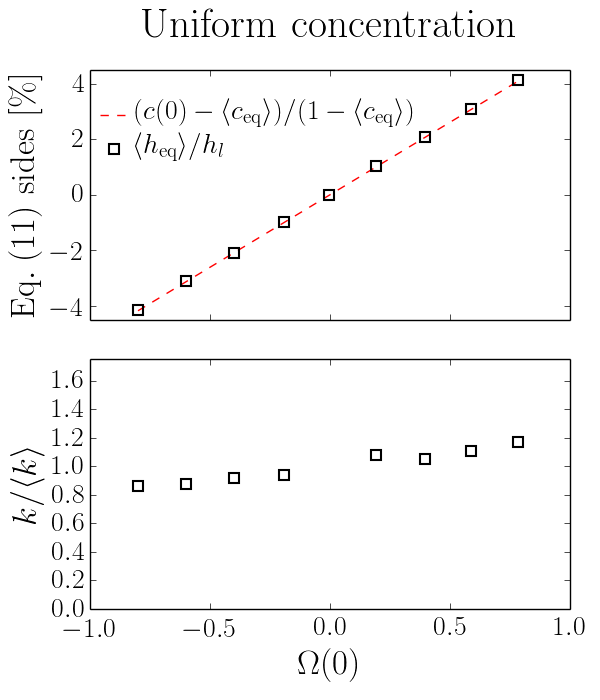} 
   \includegraphics[width=\scaleone\textwidth]{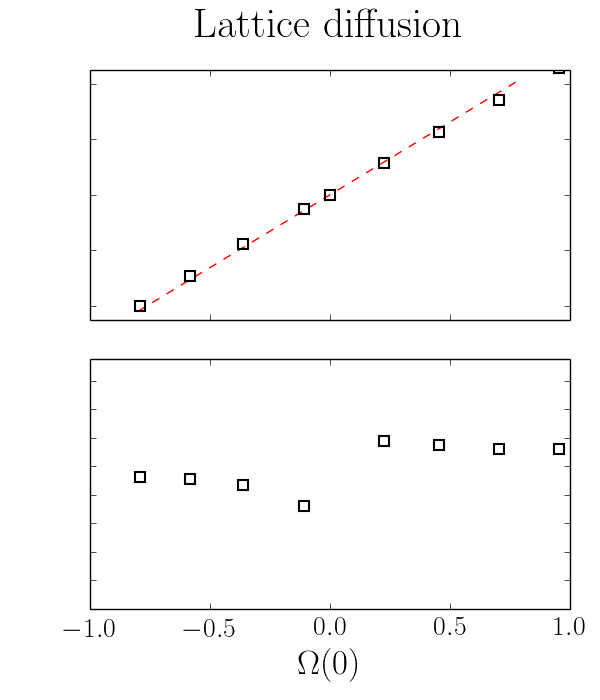} 
   \includegraphics[width=\scaleone\textwidth]{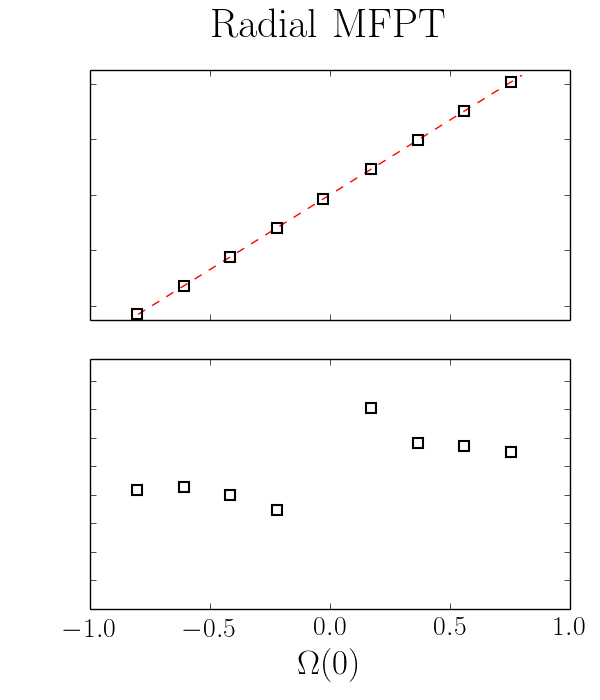} 
   \includegraphics[width=\scaleone\textwidth]{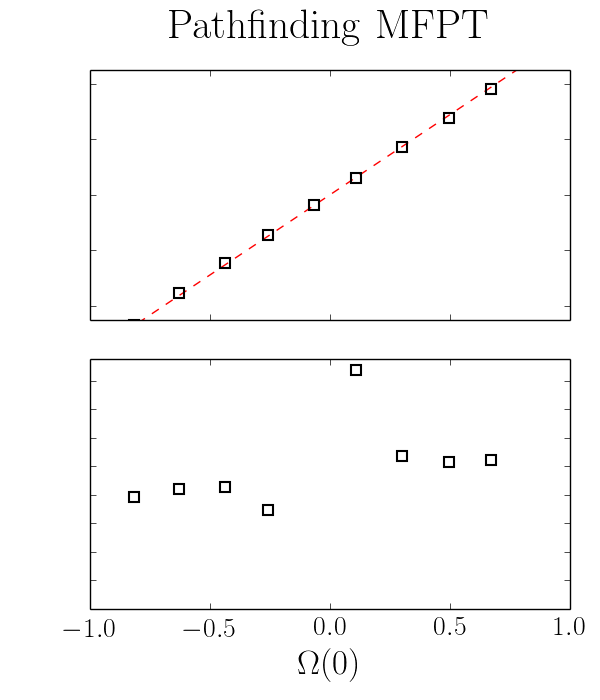} \\
   \includegraphics[width=\scaleone\textwidth]{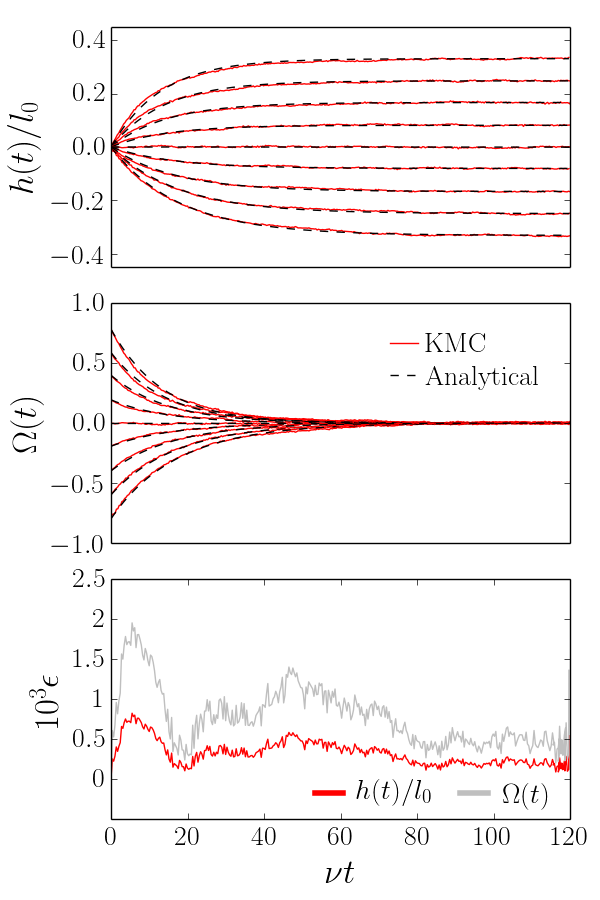} 
   \includegraphics[width=\scaleone\textwidth]{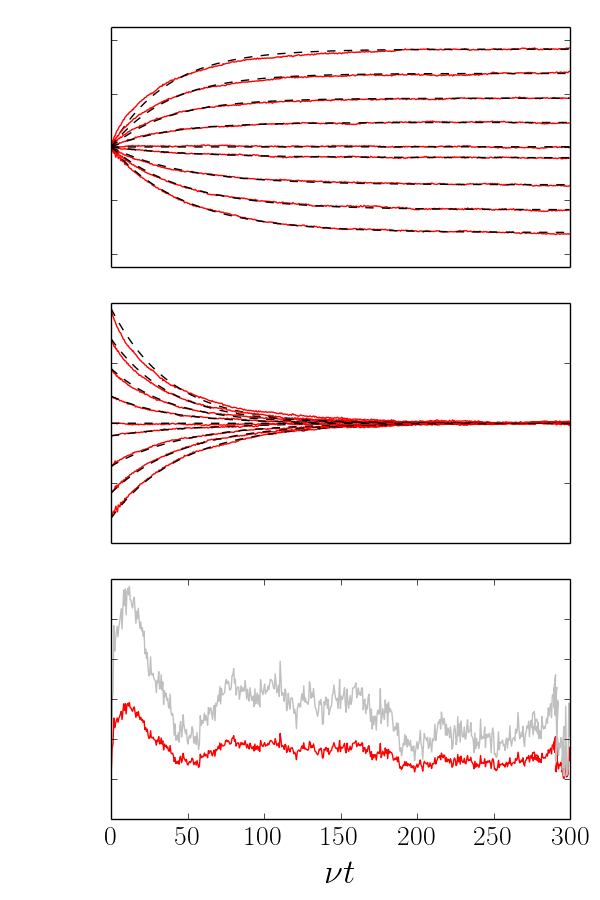} 
   \includegraphics[width=\scaleone\textwidth]{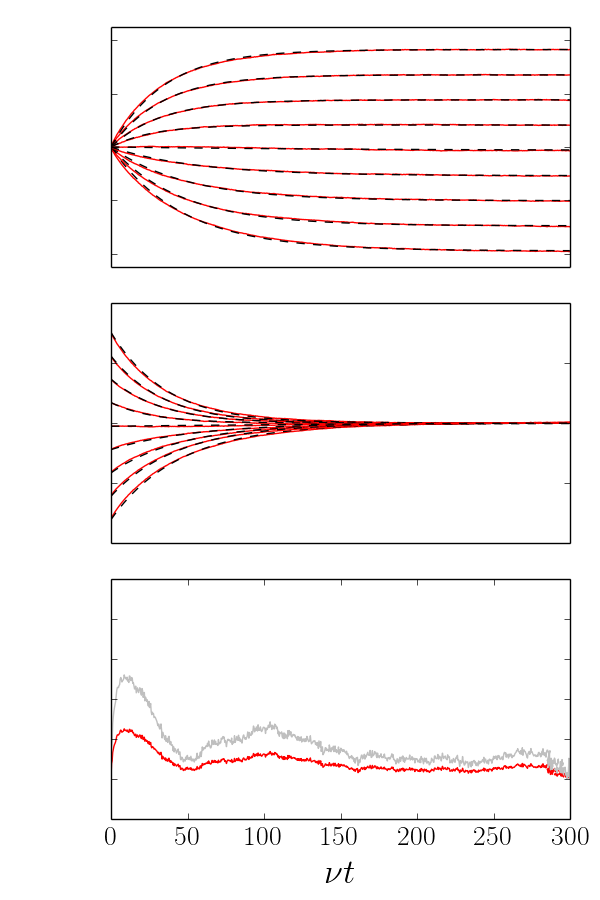} 
   \includegraphics[width=\scaleone\textwidth]{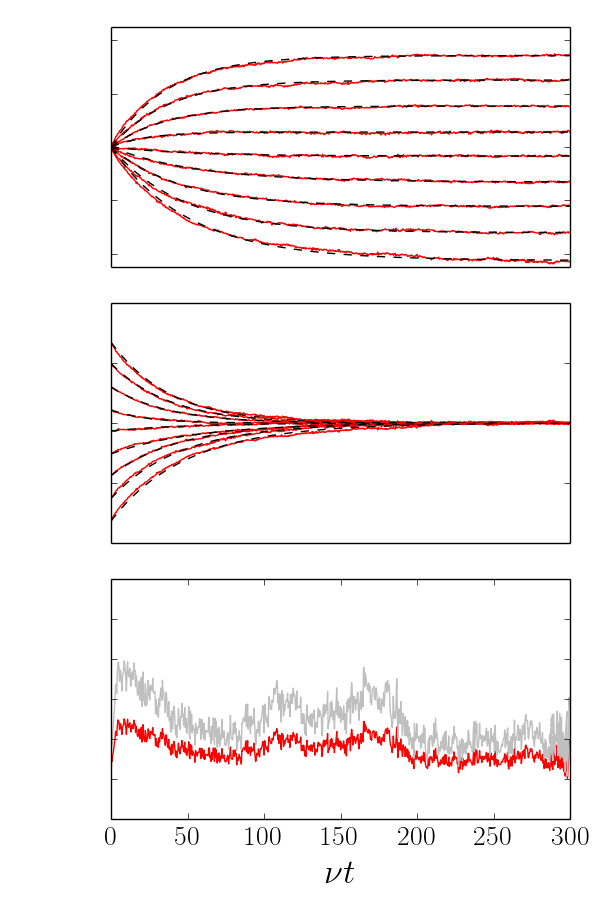}
 \caption{Comparisons between the analytical macroscopic limit and the kinetic Monte Carlo (KMC) simulations using different diffusion models (displayed
 at the top of each column). 
 The first row show the left- and right-hand side of Eq.~(\ref{eq:ht_asympt}) as squares and a stippled line, respectively, 
 which are measured independently. If these are equal, the macroscopic equilibrium condition is obeyed. As we can see this is the case.
 The second row shows the results of extracting the value of $k$ from the analytical model in \citeeq{eq:macro_dh} by least square 
 fitting of the analytical and simulation results. These $k$ are scaled by the mean value of the respective plot $\langle k \rangle$
 in order to have a unit less measure for comparing different diffusion models (which operate with different time units). 
 It is clear that there is an asymmetry between growth ($\Omega > 0$) 
 and dissolution ($\Omega < 0$), which agrees with earlier reports~\cite{Gilmer1972, Gilmer1972a, Cheng1991}.
 The remaining rows show the resulting fit of kinetic Monte Carlo (KMC) simulations and analytical results
 of the height and supersaturation as a function of time, and the average absolute error between these, respectively.
 We observe an overall good match between the analytical and simulation results.
 These results are obtained by combining 100 simulations for each initial saturation.
 The initial conditions $\Omega(0)$ for the simulations are not equal, since these are not known until the equilibrium concentration
is measured at the end of the simulation. The initial concentrations are equal for all simulations.}
\label{fig:model_results}
\end{center}
\end{figure*}

In this section we will compare the behavior of the MFPT models in the macroscopic limit
to those of the reference models.
As a system we have chosen to study one with a fixed confining surface height,
and a constant number of particles, i.e.~, we are in the conditions of the canonical ensemble.

In Appendix~\ref{app:analytical} we assume that the macroscopic equation of motion is

\begin{equation}
\label{eq:macro_dh}
 \frac{\partial h(t)}{\partial t} = k\Omega(t),
\end{equation}

\noindent
where $k$ is a constant, $h(t)$ is the average crystal surface height, and $\Omega(t) = c(t)/c_\mathrm{eq} - 1$ is the supersaturation
with $c_\mathrm{eq}$ being the equilibrium concentration, and derive analytical expressions for $h(t)$ and $\Omega(t)$,
which are given in Eq.~(\ref{eq:hexact}) and Eq.~(\ref{eq:omegaexact}), respectively.


Since the constant $k$ from Eq.~(\ref{eq:macro_dh}) is not a priori known in KMC,
a direct comparison between simulations and the analytical results from Appendix~\ref{app:analytical} is not possible.
However, the asymptotic value of $h(t)$ is

\begin{equation}
\label{eq:ht_asympt}
 \lim_{t\to\infty} h(t)/h_l = \frac{c(0) - c_\mathrm{eq}}{1-c_\mathrm{eq}},
\end{equation}

\noindent
which we see is independent of $k$ and can be directly compared. 
Moreover, since the system is in thermal equilibrium, has no external forces acting on it, and cannot exchange particles across any system boundaries, 
we know from thermodynamics that $c_\mathrm{eq} = c(t\to\infty)$.
We can therefore evaluate Eq.~(\ref{eq:ht_asympt}) using the concentration measured at a stage where it has converged $\langle c_\mathrm{eq}\rangle$, 
and compare it to the corresponding converged simulation height $\langle h_\mathrm{eq}\rangle$.
If the macroscopic equilibrium is obeyed, we should find that these two independent measurements obey \citeeq{eq:ht_asympt}.
In the top row of \citefig{fig:model_results} we see that this is the case for all the diffusion models,
where the squares and stippled lines are the left- and right-hand sides of Eq.~(\ref{eq:ht_asympt}), respectively.

Knowing that the asymptotic limits match, we now estimate $k$ by minimizing the difference between the simulated and analytical $h(t)$ and $\Omega(t)$ using least square methods.
The resulting $k$ is displayed as a function of initial saturation $\Omega(0)$ in the second row (from the top) of Fig.~\ref{fig:model_results},
and the remaining rows show the resulting simulated and analytical $h(t)$, $\Omega(t)$ and their absolute errors, respectively.
It is clear that there is an overall good match between the macroscopic limit and each of the diffusion models,
but the best match is produced by the MFPT models. It is tempting to say that this means that they are macroscopically more correct,
which may or may not be important. Nevertheless, it is remarkable that such a simple solute-surface dynamics coupling can produce 
such a high level of match with equations derived without any concern regarding these details.

The asymmetry in $k$ which has been previously observed for the uniform concentration model without confinement 
at a fixed concentration~\cite{Gilmer1972,Gilmer1972a,Cheng1991} is reproduced for all models as well, except for some 
sight diverging tendencies close to equilibrium for the free diffusion model.

These results reveal crucial properties the diffusion model should have.
For example, the results show that the equilibrium is well defined even with a direct coupling between the 
free diffusion and the lattice surface, they show that the time evolution of the system
behaves smoothly even if the diffusion is much more dynamic than in any of the other models,
and it shows that the macroscopic limit is preserved when we introduced the free diffusion model at the microscopic level.
In other words, these results show clearly that \citeeq{eq:macro_dh} is the underlying equation of the ensemble averaged motion of the surface.

We see that lattice diffusion is the model which produce the worst fit with the analytical solution.
Moreover, it seems that the height curves are not evenly spread (the gap between the equilibrium $h(t)$
and the one for $\Omega\sim -0.2$ is not identical to the other gaps). This means that the equilibrium 
concentration depends on the input concentration, which might be an effect caused by the hard-sphere 
interaction of the lattice diffusion particles.

\subsection{Equilibrium comparisons}
\label{ssec:equilibrium_comparisons}

\begin{figure*}[t]
	\begin{center}
		\includegraphics[width=0.45\textwidth]{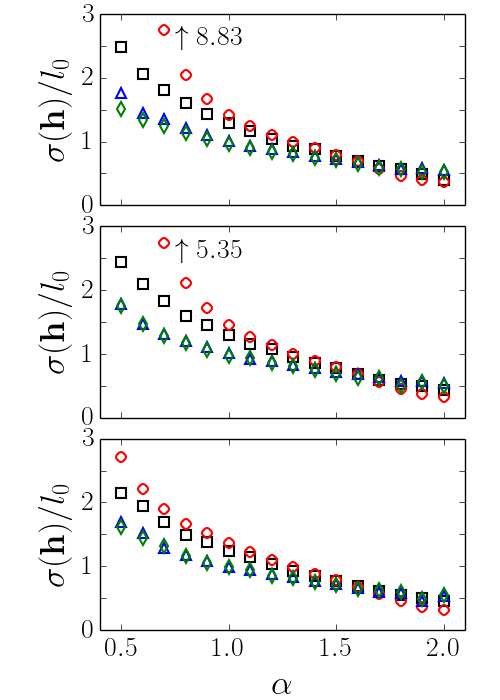} 
		\includegraphics[width=0.45\textwidth]{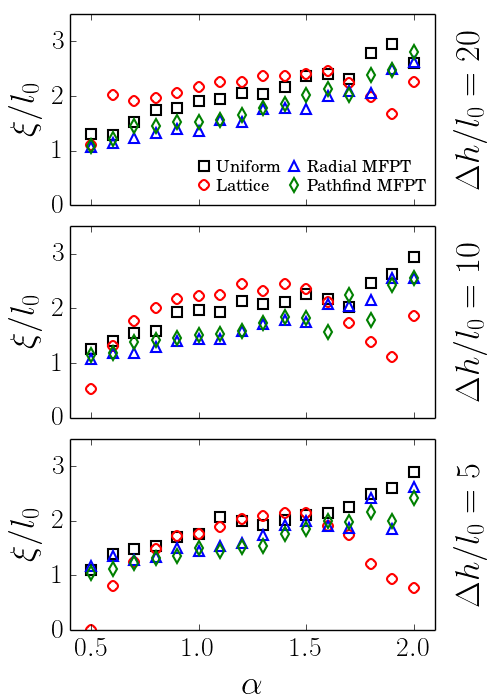} 
	\caption{Left column: The surface roughness $\sigma(\textbf{h})$ as a function of $\alpha\equiv E_b/kT$. 
	Right column: The height autocorrelation length $\xi$ as a function of $\alpha$. 
	All measurements are performed in steady state.
	The level of confinement is given by the average surface-surface separation $\Delta h$ which is held constant
	throughout the simulations. The number of solute particles are also held constant. 
	These results are obtained by averaging over 100 autocorrelation functions 
	per simulation for 20 separate simulations, except for lattice diffusion, where 1000 autocorrelation
	functions were averaged over 100 simulations. 
	We see that the roughness is decaying with increasing $\alpha$, and gets very big for 
	lattice diffusion at low $\alpha$ due to the irreversible spiking described in \citefig{fig:irreversible_lattice_diff}.
	In these cases, the value at $\alpha=0.5$ is displayed next to an arrow. 
	It is clear from these results that the free diffusion mean first passage time (MFPT) methods are well-behaved 
	as opposed to lattice diffusion, which really suffers in tight confinements.}
	\label{fig:roughness_and_correlation}
	\end{center}
\end{figure*}

\subsubsection{Roughness}

The roughness of a (2 + 1)-dimensional SOS surface described by a matrix of heights $\textbf{h}$ with elements $h_{ij}$ is calculated as 

\begin{equation}
 \sigma(\textbf{h})/l_0 = \frac{1}{A}\sqrt{\sum_{ij} \left(h_{ij} - \langle \textbf{h} \rangle\right)^2}.
\end{equation}

\noindent
and is plotted as a function of $\alpha$ for the MFPT models and the reference models in equilibrium for various separations $\Delta h$ in \citefig{fig:roughness_and_correlation}.
The overall trend is that the surface roughness decays with increasing $\alpha$, which is expected since the probability 
of dissolving a peak relative to a valley goes as $\exp(4\alpha)$.

From these results we see that the MFPT models produce similar roughnesses which generally is lower than for
the uniform concentration and lattice diffusion models. 
This tells us that there are very few scenarios where the dominant deposition path is not within radial line of sight.

For high $\alpha$ the MFPT models match the uniform concentration model,
whereas the lattice diffusion model is somewhat lower. For very low $\alpha\sim 0.5$, the simulations converge very slowly such that there 
are some errors in the results of about a quarter of a lattice length. For all other values of $\alpha$ the errors
are so small they would not be visible.

For low $\alpha$ the MFPT models do not seem to include the same diverging behavior as lattice diffusion does. 
In this limit, the equilibrium concentration is high (a lot of particles are present),
such that we expect more irreversible spiking of the kind illustrated in \citefig{fig:irreversible_lattice_diff} to occur,
which explains the sudden increase in roughness. Since the MFPT models operate with continuous coordinates,
these issues are not present. 

We see also that for higher $\alpha$, all diffusion models produce similar roughnesses, which is expected since here the
equilibrium concentration is low (few particles are present), such that the details of the diffusion model should not be important.
However, lattice diffusion produces a significantly lower roughness in tight confinement. 
This demonstrates again that the MFPT models are more well-behaved in tight confinements
than the lattice diffusion model.

As expected, the roughness becomes independent of the confinement at a certain separation where the surface roughness is no 
longer limited by the finite separation. The roughness of the MFPT methods are overall so low that 
it appears unchanged down to $\Delta h/l_0 = 5$, but the roughness of the reference models are clearly impacted
for low $\alpha$. 

\subsubsection{Height autocorrelation lengths}

The height autocorrelation function $\mathcal{C}(\delta x, \delta y)$ is calculated as

\begin{equation}
 \mathcal{C}(\delta x, \delta y) = \langle (h_{xy} - \langle \textbf{h}\rangle)(h_{(x + \delta x)(y + \delta y)} - \langle \textbf{h}\rangle) \rangle,
\end{equation}

\noindent
where the average is over all lattice sites $(x, y)$. It is often custom to scale this function such that $\mathcal{C}(0.0) = 1$,
however, since we are interested in the autocorrelation length $\xi$, this is not necessary. 
The autocorrelation length $\xi$ is defined as the inverse decay constant of the autocorrelation function, that is, if

\begin{equation}
\label{eq:correlation_function}
\mathcal{C}(\delta x, \delta y) \sim \exp(-\sqrt{\delta x^2 + \delta y^2}/\xi),
\end{equation}

\noindent
then we can extract $\xi$ from a curve-fit of the sampled autocorrelation function.

We have calculated $\xi$ using four directions: along the $y$-axis for $\delta x = 0$ and opposite,
and for the two diagonals going through $\mathcal{C}(0, 0)$. In all models we
found the autocorrelation length of these directions to match up, and therefore we
calculated $\xi$ as the average of these four values.
The results as a function of $\alpha$ for all diffusion models in equilibrium for various separations are given in \citefig{fig:roughness_and_correlation}.

The trend is that the autocorrelation length increases with $\alpha$.
This makes sense since as $\alpha$ increase, 
the system is driven more and more towards minimizing the surface tension,
which means that if one surface site is at a certain height, it becomes more likely that the surroundings are at a similar height,
i.e.~we increase the autocorrelation length.

Again, as for the roughness, the MFPT models seem to produce similar results.
This again suggests that the dominant deposition sites are likely to be in line of sight.
We see also that the MFPT autocorrelation lengths are generally lower than those of the reference diffusion models,
except for $\alpha < 0.75$ and $\alpha>1.5$, where the lattice diffusion model behaves chaotic.
Looking at the respective roughness plot in the same figure we see that these points
are the same as where the roughness diverges and drops below that of the uniform concentration model.
Hence this further demonstrates that lattice diffusion struggles in tight confinement, whereas
the MFPT models are generally well-behaved.

The chaotic behavior for lattice diffusion is found to be due to the autocorrelation function not resembling an exponential decay,
hence we are unable to extract $\xi$ from \citeeq{eq:correlation_function}
in a well-defined manner.

It is interesting to note that all diffusion models produce autocorrelation lengths 
which has the same slopes with increasing $\alpha$.

\subsection{Example system: Cavity with flux boundary}
\label{sec:example}

\newcommand{\wfac}{0.8}
\begin{figure*}[t]
	\begin{center}
	\includegraphics[width=0.83\textwidth]{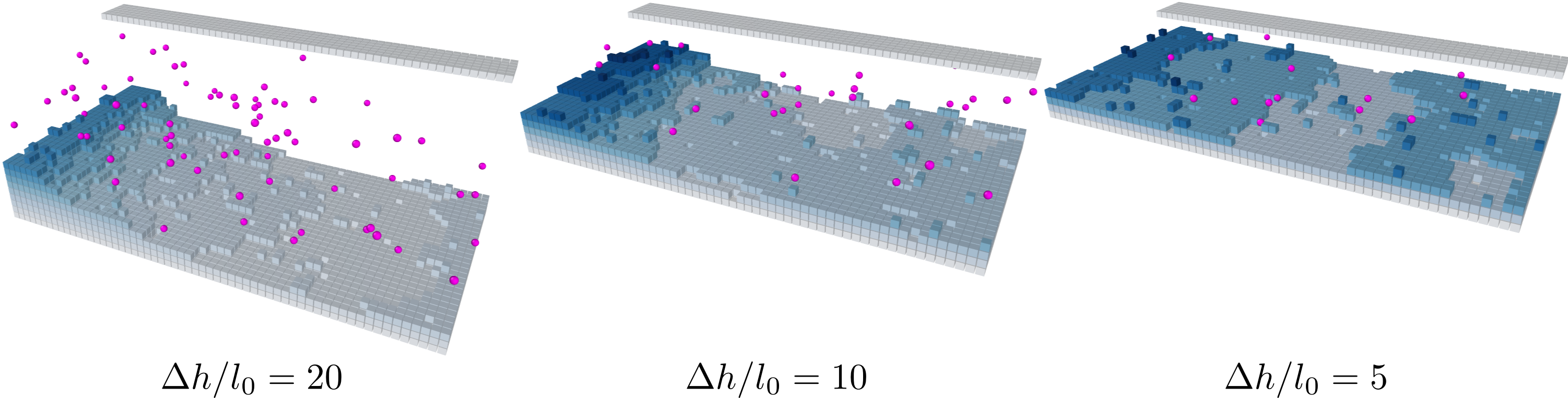} 
	\caption{Snapshots taken at various levels of confinement $\Delta h$
	from the system described in Sec.~\ref{sec:example} with a reflecting right boundary. 
	Only the first five cells of the confining surface (top) is rendered to better view the surface (cubic particles)
	and the solute particles (spherical). The surface color intensity is proportional to the vertical position.
	The images are rendered using Ovito~\cite{Stukowski2009a}.}
	\label{fig:3dsurfs}
	\end{center}

	\begin{center}
	\includegraphics[width=\wfac\columnwidth]{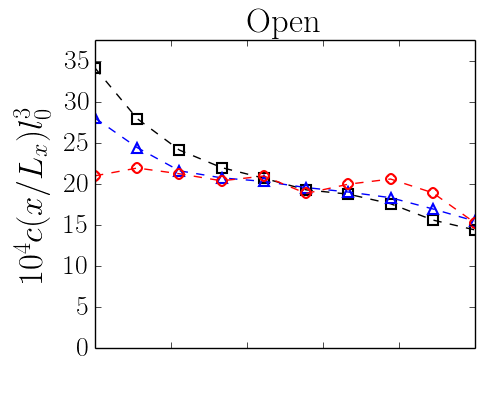} 
	\includegraphics[width=\wfac\columnwidth]{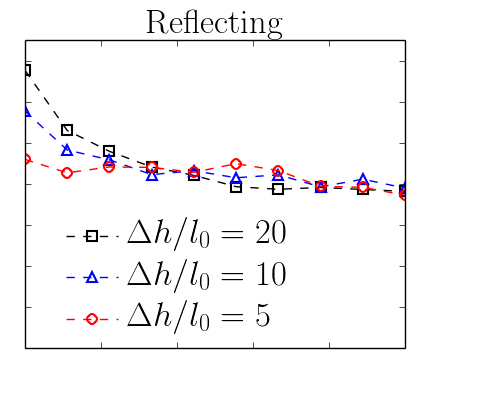} \\
	\includegraphics[width=\wfac\columnwidth]{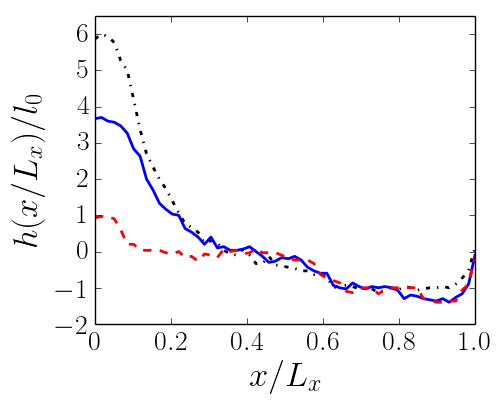} 
	\includegraphics[width=\wfac\columnwidth]{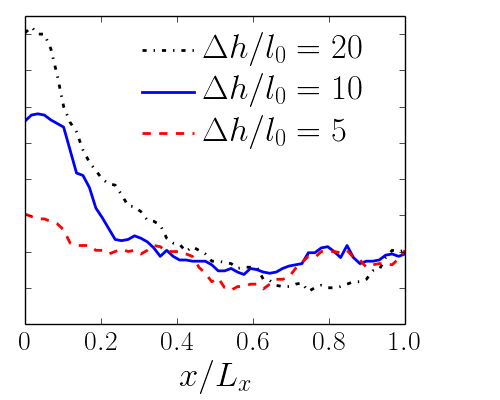} 
	\caption{Concentration and height profiles for surfaces in various levels of confinement $\Delta h$ with 
	a flux boundary condition $J_\mathrm{in}=3\nu/l_0^2$ on $x=0$ and a different boundary condition on $x=L_x-1$, where $L_x=60$ is the system length,
	indicated as the respective column title (open or reflective). The open boundary has a reverse flux of $J_\mathrm{open} = \nu/l_0^2$ to keep the surface from 
	dissolving indefinitely. The surface is pinned meaning that particles at the $x$-boundaries with $h_{ij}\le0$ has an extra neighbor.
	Periodic boundary conditions are applied in the $y$-direction where $L_y=30$. 
	The concentration $c(x)$ is measured as the number of particles per volume using ten bins across the domain and is time averaged.
	A similar analysis for the $y$-component of the concentration has been done to ensure that it is uniform.
	The height profiles $h(x)$ are the $y$-averages of the surfaces at the end of the simulations.
	The surfaces are shown in three dimensions in Fig.~\ref{fig:3dsurfs} for the reflecting boundaries.
	The surfaces are more stable near a reflecting boundary than an open one, since solute particles cannot escape in the former case.
	The concentration profile from the point where the steps end decay linearly for the open boundary, however, in the region
	where the steps are present, i.e.~the region of growth, the decay is non-linear.
	For the reflecting boundary we get a smoother decay since the boundary requires $\partial c(x)/\partial x=0$ at $x=L_x-1$.
	The exception to these observations is the $\Delta h/l_0=5$ case, where the concentration field decays differently.
	In this case, there are only three vertical cells in the liquid that are away from the surface,
	which makes it difficult to measure the concentration in a well-defined manner.}
	\label{fig:diff_comp}
	\end{center}
\end{figure*}

We have up to this point focused on showing that the MFPT models behave well in confinement and in the macroscopic limit. 
Here we will apply the radial MFPT model to a system where the reference models would not be suited.
The system we will study is the same as the periodic one used in the macroscopic limit, however, now with the 
$x=0$ boundary opened up (solute particles can escape) with an added flux boundary condition.
For the opposing boundary at $x=L_x-1$ we study two different boundary conditions: open with a lower flux, and reflecting.
The surface is also pinned in $x=0$ and $x=L_x-1$ such that a boundary site has no additional neighbor if its height $h_{ij} > 0$.

We add a flux $J_\mathrm{in}=3\nu/l_0^2$ at $x=0$ and a flux $J_\mathrm{open}=\nu/l_0^2$ at $x=L_x-1$,
which is translated into rates by multiplying by the open boundary area. The reflecting boundary simply mirrors the particle 
across the boundary when it diffuses outside. Periodic boundaries are still applied in the $y$-direction.
Snapshots of the system at different levels of confinement are given in Fig.~\ref{fig:3dsurfs}.

From the results in Fig.~\ref{fig:diff_comp} we see that the crystal surface height averaged over the $y$-direction, $h(x)$,
consist of a number of steps which is at its highest close to the boundary and fluctuates between partial or complete pits (negative heights) and 
partial or no islands deeper into the cavity. 
The reason why the decay is relatively smooth is that the step fronts fluctuate. 
The bunched steps forms a growth rim such as is observed experimentally for confined crystals growing in 
a supersaturated solution~\cite{G.F.Becker1916, Taber1916, Royne2012}. Close to the inner boundary we get either a pit due to the low flux and open boundary,
or a stable evolution from the middle of the channel due to the reflecting boundary.

The concentration field $c(x)$ have been measured as the average number of particles per volume.
We have studied the dependency in the periodic direction and found it to be uniform, hence $c(x)$ is averaged over the $y$-direction.
Figure~\ref{fig:diff_comp} show that for the open boundary, the concentration field is linearly decaying from the point where the nucleated steps have ended,
and has a non-linear dependency close to the high flux boundary where steps nucleate. For $\Delta h/l_0=5$ there is only three vertical cells 
which are away from the surface initially, which makes it hard to calculate the concentration field in a well-defined manner. 
Hence the concentration field here is not as well-behaved. This fact demonstrates the necessity of the discrete representation of the particles.
For the reflecting boundary, the concentration field curves off such that $\partial c(x)/\partial x=0$ at the inner boundary.

These results demonstrate that the MFPT diffusion model is well suited for simulation in these environments,
and is able to produce the correct diffusion dynamics as well as the correct surface dynamics with a direct coupling between the two. 
The results also demonstrate that the produced concentration field responds correctly to other boundary conditions than the periodic ones used
in the previous sections. A uniform concentration model is not applicable in environments such as the one presented here,
and guessing the shape of the concentration field is not trivial. It is therefore clear that in order to model these
systems in a satisfactory manner, a discrete particle free diffusion model which explicitly solves the diffusion equation is indeed necessary.

\section{Discussions and Conclusions}
\label{sec:Discussions}


We have presented an algorithm for coupling free particle diffusion in a liquid to lattice crystal surface models subject to spatial confinement,
where the dynamics is described in terms of rates of single particle events (dissolution and deposition).
The basic idea behind the algorithm is that for every solute particle, a deposition rate is calculated 
to sites present inside a cube centered around the particle with sides $2r_\mathrm{max}+1$. 
This rate is the inverse of a mean first passage time (MFPT) of depositing at the site,
which is found to scale as $r^4$, where $r$ is a characteristic distance to the site.
This characteristic distance was modeled in two ways: radially and as the shortest available path (accounting for line of sight).
We have shown that the model excels at simulating confined geometries in which more basic existing models are not well suited.

In this paper we have used a (2+1)-dimensional solid-on-solid (SOS) surface on a cubic lattice, but the derived diffusion model
is not limited to this. In practice the model should be directly applicable to any surface with predefined deposition sites. 
The strength of the diffusion model derived here is that it functions well in all types of geometries, including spatial confinement. 
However, the diffusion model should work just as well for open geometries (e.g.~a single surface in an infinite bath).   

The radial and pathfinding MFPT models were found to produce a similar equilibrium roughness, a similar equilibrium height autocorrelation length,
and both produced the correct macroscopic limits. For growth, however, there was an increasing difference between the two
models, since the radial model ignores line of sight issues. Unless the growth is very rapid, we
found that the two methods produce near identical dynamics, which means that the most likely 
deposition sites are in line of sight of the particle. This makes sense since the $r^{-4}$ term decays
so quickly that a particle is most likely very close to the most dominant deposition site.
Based on this it is fair to say that it is safe to use the radial method except when the system is
growing very rapidly. This also saves a lot of CPU time, since the pathfinding is very slow compared to 
calculating a straight line.

Besides producing the correct macroscopic limit, 
the free diffusion also produce a known asymmetry in the kinetic constant $k$ which appear in the macroscopic equation of motion. 
The free particle model have a slight divergence in $k$ close to equilibrium, however, 
this means only that the assumption that the speed is proportional to the supersaturation to unit power is false close to equilibrium.
This non-linearity is observed experimentally close to equilibrium~\cite{Khorsand2010,Bennema1967a}.
We will in other words not only automatically obtain the correct concentration field, but we 
include non-linear effects which a uniform concentration model does not.



It was clear that the lattice diffusion method had several issues in tight confinements,
none of which were present in the MFPT models. The lattice diffusion is less CPU intensive per cycle than 
the free diffusion, however, in the free diffusion we move all the particles at once 
while we in the lattice diffusion model move the particles one at the time.
Hence if we look at the CPU time required per surface event, the free diffusion 
is more efficient. 

The cavity example demonstrated that the MFPT model reacts well to different boundary conditions, and that it 
leads to a correct surface evolution even in systems which are not infinite and isotropic. Moreover, 
it demonstrated a challenge with measuring the concentration in a well-defined manner close to the surface.

With a free diffusion model available it should be possible to include for instance a drift term into the diffusion equation
assuming that the MFPT is still dominated by the diffusion. 
It should also be possible to force the solute particles to follow a specific distribution using a Metropolis or Metropolis Hastings Monte Carlo algorithm.
Including solute particle chemistry and interactions in general through molecular dynamics or Boltzmann statistics and the Metropolis algorithm should also be possible, 
however, this means that the MFPT expression should depend on the details of these additions as well. 

\begin{acknowledgments}
  Discussions and comments by Anja R{\o}yne significantly contributed to this paper. 
  This study was supported by the Research Council of Norway through the project ``Nanoconfined crystal growth and dissolution'' (No. 222386). 
  We acknowledge support from the Norwegian High Performance Computing (NOTUR) network through the grant of machine access.
\end{acknowledgments}

\appendix

\section{Lattice Diffusion Model}
\label{app:lattice}

If we restrict the motion of solute particles to a lattice which is an extension of the crystal lattice into the liquid,
we get the diffusion model we refer to as lattice diffusion. Since the liquid is assumed to be ideal, 
solute particles do not bind to one another, however,
hard-sphere interactions are unavoidable since two particles cannot occupy the same cell at the same time.
Inserting $E_b=0$ into Eq.~(\ref{eq:rate}) yields $\mathcal{R}(p_s) = \nu$ for all solute particle positions $p_s$, 
which is the rate of a solute particle transitioning 
away from its current location. This model is illustrated in the middle part of \citefig{fig:diffusion_models}.

Solute particles transitioning to the lattice sites immediately above the surface instantly become surface particles. 
This constraint guarantees that the layer immediately above the surface is always vacant,
hence it should not be included when calculating the concentration. The concentration becomes

\begin{equation}
\label{eq:c_lattice_orig}
 c(t) = N(t)/V_f(t),
\end{equation}

\noindent
where $N(t)$ is the number of solute particles, and $V_f(t) \equiv V(t) - Al_0$ is the volume at which solute particles are free to move, 
where $V(t)$ is the cavity volume and $A$ is the reactive surface area.

\subsubsection*{Deposition rates}

Since we do not distinguish between adsorbed and free surface particles, 
a deposition is simply a solute particle transitioning from a site in the liquid to a site on the surface. 

This means that there is no need for an explicit deposition reaction (and rate), since 
it is implicit in the lattice diffusion model. However, for the sake of completeness
we may write the deposition rate at surface site $i$ as

\begin{equation}
\label{eq:deprate_lattice}
 \mathcal{R}_+^{(l)}(i) = \sum_{\langle ij\rangle} \sigma(j) \mathcal{R}(p_s^{(j)}) = \nu \sum_{\langle ij\rangle} \sigma(j),
\end{equation}

\noindent
where $\langle ij\rangle$ denotes a sum over liquid sites $j$ neighboring site $i$, and $\sigma(j)$ is 1 if a particle is present at liquid site $j$, and 0 otherwise. 

\subsubsection*{Implementation details}

In this model, the rate $\mathcal{R}(p)$ of a particle $p$ transitioning away from a current position is independent of the destination.
This means that if the particle has $n_-$ escape paths, the total escape rate becomes $n_-\mathcal{R}(p)$. 
If the particle escape is selected, the destination is then selected at random. Alternatively we could
add $n_-$ reactions to the catalog with different escape paths.

The rates of transitions in the liquid are larger than those of the surface, and if the concentration is high, the number of
transitions available for solute particles is large compared to for surface particles. This means
that we spend a lot of time moving the solute particles. However, whenever we move a particle,
we only have to recalculate the rates of the affected sites, which on average is very few.
Nevertheless, the efficiency of this model suffers greatly from the constraint of moving a single particle at the time.

If a particle from the liquid deposits by transitioning sideways onto a surface site such that it immediately connects one or more particles
directly above this site to the surface, the surface height can increase more than one unit per step in lattice diffusion.
However, since a particle has to dissolve from the top site due to the solid-on-solid (SOS) condition, 
the height cannot decrease by more than one unit per step.
This means that all transitions are not completely reversible,
and creates a tendency of the system ``spiking'' at high concentrations.
This process is displayed in \citefig{fig:irreversible_lattice_diff}.
It is possible to avoid these issues by for instance repositioning of the particles causing the issues,
however, our focus is not on creating ad hoc fixes to problematic models,
but rather unveiling what is causing the problems and study how they limit
one model as compared to another.

\begin{figure}
	\begin{center}
		\includegraphics[width=0.45\textwidth]{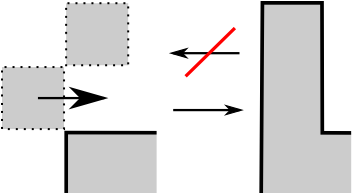}
	\caption{Illustration showing an irreversible transition in lattice diffusion growth on a solid-on-solid (SOS) surface (solid outline). 
	A solute particle (stippled outline) depositing sideways onto the surface (thick arrow in left figure) may connect several particles to the surface
	by doing so (resulting in the right figure). However, the reverse is not possible in a single transition since only the top particle may dissolve in a SOS model.
	This leads to a ``spiking'' tendency.
	}
	\label{fig:irreversible_lattice_diff}
	\end{center}
\end{figure}

\section{Uniform Concentration Model}
\label{app:uniform}

If we simply average over the liquid sites in the lattice diffusion model and describe it by a uniform concentration $c(t)$,
we get what we refer to as the uniform concentration model. As we will see, this model will correspond to a constant (in space) deposition rate model,
also referred to as a random deposition model. This model is displayed in the right-most figure of \citefig{fig:diffusion_models}.
What we effectively do is to only count the number of particles in the liquid, not caring about where they are located.

By rearranging \citeeq{eq:c_lattice_orig} we can calculate the initial number of particles at $t=0$ as $N(0) = c(0)V_f(0)$.
We can track the change in solute particles $\Delta N(t)$ by counting the number of depositions and dissolutions,
such that $N(t) = N(0) + \Delta N(t)$. We get

\begin{equation}
\label{eq:c_uniform}
 c(t) = \frac{N(t)}{V_f(t)} = \frac{N(0) + \Delta N(t)}{V_f(t)} = \frac{c(0)V_f(0) + \Delta N(t)}{V_f(t)}.
\end{equation}


\subsubsection*{Deposition rates}

By averaging \citeeq{eq:deprate_lattice} over all surface sites $i$, 
we can calculate the deposition rate for the uniform concentration model by 
setting this average occupancy equal to the uniform concentration $c(t)$ as follows:

\begin{equation}
 \mathcal{R}_+^{(c)} = \langle \mathcal{R}_+^{(l)}(i)\rangle = n_+ \nu \langle \sigma \rangle = n_+ \nu c(t),
\end{equation}

\noindent
where $n_+$ is the number of surrounding liquid sites.
This is true only if the concentration is uniform, since in general the concentration might be higher or lower close to the surface than other places (e.g.~in bulk).

\subsubsection*{Implementation details}

When a particle is deposited, we do not know from which one of the $n_+$ liquid sites it came. However,
we know that the number of dissolution paths $n_-$ necessary equals $n_+$ for the immediate reverse transition.
It can be shown that in steady state, such factors that are independent of the final state, and is equal
for a transition and its immediate reverse transition, can be removed from the rates.
We therefore do not count paths in the uniform concentration model.

Since we do not track the solute particles explicitly there is nothing stopping the concentration from
becoming negative. In order to avoid this, we set the concentration to zero in the cases where the value of
$\Delta N(t)$ is so that $N(t)$ would become negative.

Without confinement, the equilibrium concentration is known to be~\cite{Gilmer1972}

\begin{equation}
\label{eq:c_eq_analytical}
 c_\mathrm{eq} = \exp(-3\alpha).
\end{equation}

\section{Analytical Calculations for the Macroscopic Limit}
\label{app:analytical}

%
%
%
Crystal growth and dissolution with fast surface integration is empirically known to obey the following relationship~\cite{Machej1997,Khorsand2010,Bennema1967a,Haneveld1971,Dove2005}

\begin{equation}
\label{eq:dh_omega}
\frac{\partial h(t)}{\partial t} = k\Omega(t),
\end{equation}

\noindent
where $k$ is a constant, $h(t)$ is the average height of the surface, and $\Omega(t) = (c(t)-c_\mathrm{eq})/c_\mathrm{eq}$ measures the degree at which the system is out of equilibrium.

Here we will solve this equation for $h(t)$ and $\Omega(t)$ 
for a closed system which can only exchange heat with its surroundings, 
i.e.~in the conditions of the canonical ensemble.

Let the position of the confining surface be $h_l$, then the cavity volume $V(t)$ is

\begin{equation}
\label{eq:v_t}
 V(t) = (h_l - h(t))A = V_s + N(t)l_0^3,
\end{equation}

\noindent
where $A$ is the reactive surface area, $V_s$ is the volume of the solvent, 
and $N(t)l_0^3$ is the volume of the number of particles $N(t)$.
Since the system is closed and a particle takes up an equal amount of space in the liquid as on the surface, 
the volume of the solvent $V_s$ is constant. We calculate $V_s$ from the initial conditions as 

\begin{align}
 V_s &= V(0) - N(0) = V(0)(1 - c(0)) \notag \\
     &= A(h_l - h(0))(1 - c(0)),
\end{align}

\noindent
where we used that the unit less concentration is $c(t) = l_0^3N(t)/V(t)$.

By using \citeeq{eq:v_t} we can express $c(t)$ in terms of the height

\begin{align}
 c(t) = l_0^3\frac{N(t)}{V(t)} &= \frac{V(t)-V_s}{V(t)} = 1 - \frac{V_s}{V(t)}\notag \\
                               &= 1 - \frac{V_s}{A(h_l-h(t))},
\end{align}

\noindent
such that the supersaturation $\Omega(t)$ becomes

\begin{equation}
\label{eq:omegat}
\Omega(t) = \left(\frac{1}{c_\mathrm{eq}} - 1\right) - \frac{V_s}{Ac_\mathrm{eq}}\frac{1}{h_l - h(t)}.
\end{equation}

Defining $K \equiv kV_s/Ac_\mathrm{eq}$ and $C\equiv k(1/c_\mathrm{eq} - 1)$, we get by inserting 
\citeeq{eq:omegat} into \citeeq{eq:dh_omega} that

\begin{equation}
\label{eq:h_exact_firstderiv}
 \frac{\partial h(t)}{\partial t} = C - K\frac{1}{h_l - h(t)},
\end{equation}

\noindent
whose solution is

\begin{equation}
\label{eq:hexact}
 h(t) = h_l - \frac{K}{C}\left(1 + W(\xi \exp(-C^2t/K))\right),
\end{equation}

\noindent
where $W(z)$ is Lambert W-function, that is, the inverse function of $f(z) = z\exp(z)$.

Inserting \citeeq{eq:hexact} into \citeeq{eq:h_exact_firstderiv}, we can obtain the expression for $\Omega(t)$ by rearranging \citeeq{eq:dh_omega}. We get

\begin{equation}
\label{eq:omegaexact}
\Omega(t) = \frac{C}{k}\left(1 - 1/\left(1 + W(\xi \exp(-C^2t/K))\right)\right),
\end{equation}

\noindent
where the constant $\xi$ is expressed in terms of the initial condition $\Omega(0)$
as 

\begin{equation}
 \xi = \left(\frac{c_\mathrm{eq}\Omega(0)}{1 - c_\mathrm{eq}(1 + \Omega(0))}\right)\exp\left(\frac{c_\mathrm{eq}\Omega(0)}{1 - c_\mathrm{eq}(1 + \Omega(0))}\right).
 \end{equation}

%
%
%
%

\bibliography{paper2_refs}
\bibliographystyle{apsrev4-1}

\end{document}